# Dynamic Resource Provisioning of a Scalable E2E Network Slicing Orchestration System

Ibrahim Afolabi, Jonathan Prados-Garzon, Miloud Bagaa, Tarik Taleb, and Pablo Ameigeiras

***Abstract*—Network slicing allows different applications and network services to be deployed on virtualized resources running on a common underlying physical infrastructure. Developing a scalable system for the orchestration of end-to-end (E2E) mobile network slices requires careful planning and very reliable algorithms. In this paper, we propose a novel E2E Network Slicing Orchestration System (NSOS) and a Dynamic Auto-Scaling Algorithm (DASA) for it. Our NSOS relies strongly on the foundation of a hierarchical architecture that incorporates dedicated entities per domain to manage every segment of the mobile network from the access, to the transport and core network part for a scalable orchestration of federated network slices. The DASA enables the NSOS to autonomously adapt its resources to changes in the demand for slice orchestration requests (SORs) while enforcing a given mean overall time taken by the NSOS to process any SOR. The proposed DASA includes both proactive and reactive resource provisioning techniques). The proposed resource dimensioning heuristic algorithm of the DASA is based on a queuing model for the NSOS, which consists of an open network of G/G/m queues. Finally, we validate the proper operation and evaluate the performance of our DASA solution for the NSOS by means of system-level simulations.**

## I. Introduction

The Network Slicing concept plays a significant role in Fifth Generation (5G) mobile networks. Network slicing enables network operators to dynamically create logical and isolated network partitions, dubbed network slices, to deliver customized network services for the different market use cases, which demand diverse and sometimes opposing requirements, on top of a common physical network infrastructure [2], [3]. Most importantly, the network slices are orchestrated to facilitate the business solutions of mostly 5G vertical industries/services and over-the-top application providers. The fundamental enabling technologies of Network slicing are Network Functions Virtualization (NFV) [5]–[7], which decouples the different network functions from the underlying hardware, and Software-Defined Networking (SDN) [8] that separates the control and forwarding planes. These technologies allow network virtualization, i.e., the ability to provide a logical software-based view of the hardware and software networking resources, in order to create virtual networks (or network slices) that are independent on the substrate network hardware.

Network slicing leverages the combination of network virtualization and infrastructure-as-a-service paradigms to enable the automation of key network slices management operations such as deployment and scaling. In this way, operators will be able to handle workload fluctuations with great agility and in a cost-effective way, while keeping the desired performance for the different network slices.

Despite the network slicing benefits, the end-to-end (E2E) orchestration and management of the different network slices can be a difficult endeavor [9]. The main tasks of a Network Slicing Orchestration System are [9]: i) network slices creation based on slices requirements upon the physical network, ii) virtual network embedding (VNE) and function placement [40]–[43], and iii) lifecycle management of the network slices including their dynamic resource provisioning (DRP) or scaling. To accomplish these tasks, the NSOS will have to run complex and computationally intensive optimization algorithms. Furthermore, the NSOS might orchestrate simultaneously a large number of network slices given the plethora of vertical services envisaged for 5G. On top of this, network slices for critical applications will provide service to a narrow area due to their ultra-low latency requirements, thus increasing the number of network slices to manage. All these network slices might generate a large number of slice orchestration requests (SORs) (e.g., slice creation, release or scaling queries) that eventually saturates the capacity of the NSOS to serve those slice requests. By way of illustration, the reactive provisioning of an Internet application can be triggered once every few minutes [10], then it can potentially generate dozens of scaling requests per hour. It shall be noted, however, that even in scenarios with low SORs arrival rate the demand of computational resources might fluctuate considerably due to the high service times at some NSOS components.

In the light of the above-described scenario, there is a need to tackle the DRP problem of the NSOS to face workload fluctuations without human intervention [11], while assuring an efficient utilization of the resources allocated to it. In other words, algorithms that will enable the NSOS to decide when, how and how much of resources to be provisioned itself. The main challenge of the NSOS dynamic provisioning is to perform the sizing of a high number of entities (refer to Section III-B) at once and as quickly and efficiently as possible. The DRP problem has been addressed in the literature mostly for internet web applications [25] and softwarized network

Ibrahim Afolabi, Miloud Bagaa and Tarik Taleb are with the Department of Communications and Networking, School of Electrical Engineering, Aalto University, Finland (e-mails: {firstname.lastname}@aalto.fi). Tarik Taleb is also with the Department of Computer and Information Security, Sejong University, Seoul, South Korea. Jonathan Prados-Garzon is with the Research Centre for Information and Communications Technologies of the University of Granada (CITIC-UGR) and Department of Signal Theory, Telematics and Communications, University of Granada, Granada, Spain (e-mails: jpg@ugr.es).

services [28], [29], [31], [32], [35]. However, to the best of the authors knowledge, the already proposed DRP solutions either are not scenario-agnostic or cannot offer simultaneously low-complexity, joint sizing of all entities, and system stability (refer to Section II).

In this paper, we propose a novel holistic global E2E mobile NSOS that enables network slicing for the next generation mobile networks by specifically considering all the aspects of the mobile network spanning across the access, core and transport parts. Our presented high-level architecture shows a hierarchical E2E network slicing system composed of a global orchestrator (GO) and multiple domain-specific orchestrators (DSOs) and their respective system components. Although, different levels of management exposure is possible when orchestrating a network slice as in [4], however, our focus is on the one that allows the customer to request slice orchestration and monitor them only. Moreover, we propose and develop a novel Dynamic Auto-Scaling Algorithm (DASA) for the NSOS that would instantaneously react to changes in the orchestration system's workload. By so doing the overall E2E NSOS's response time to serve network slices orchestration requests (NSORs) within a given target delay is maintained. The NSOS's response time is understood here as the sum of all the processing and waiting times experienced by a SOR when passing through the different NSOS's entities during its lifetime in the NSOS. To the best of the authors' knowledge, this is the first work in which a federated slice orchestration system is proposed and its dynamic resource provisioning (DRP) addressed.

The proposed DASA includes both proactive and reactive provisioning mechanisms. The proactive mechanism relies on a workload predictor, which is implemented using machine learning techniques. The reactive provisioning module is responsible for triggering asynchronous requests to scale in or out the different entities of the NSOS. The entities of the system are later detailed in Section III-B. The core of the whole solution is a resource dymensioning heuristic algorithm which is in charge to determine the required amount of computational and virtual resources to be allocated to the NSOS for a given workload so that a maximum response time of the NSOS is guaranteed. Namely, the resource dimensioning algorithm will be invoked when a provisioning decision is taken to decide how much resources have to be requested or released. Throughout this article, we will consider that the different NSOS entities run CPU-bound processes, though our solution can be easily adapted to more general scenarios. Under this consideration, the heuristic algorithm provides the required number of physical CPU cores and virtualization containers to be allocated to each NSOS entity. To that end, the heuristic algorithm relies on the use of a performance model to estimate the response time of the NSOS. Consequently, in this work, we develop a holistic performance model of the NSOS using queuing theory. The model follows the same modeling approach as in [23]. More precisely, the NSOS is modeled as an open network of G/G/m queues. The resulting network of queues is solved using the Queuing Network Analyzer (QNA) method [22].

Our DRP solution interacts with the request policing mechanism of the NSOS which is responsible for rejecting excess requests when either there is unforeseen workload surges or it becomes impossible to scale the system due to lack of resources. That is because of these two modules are closely coupled, since the DRP mechanism provides the maximum workload peak that the system could withstand at a given point, which helps to configure the request policing part. In the same way, the admission control part could invoke the provisioning mechanism when the request drop rate exceeds a certain threshold.

We validate the proper operation and evaluate the performance of the proposed DRP solution by means of system-level simulations. The results show that, for the scenario considered, the algorithm is able to find the minimal required resources to keep the mean response time of the NSOS under a given threshold. Our results also suggest that the NSORs rejection rate during a given period, is determined by the reaction time of the reactive provisioning mechanism which, in turn, strongly depends on the Virtual Machine (VM) instantiation time.

The remainder of the paper is organized as follows. Section II briefly reviews the related literature. Section III describes in detail the proposed network slicing orchestration system for the next generation mobile networks. Section IV includes the system model and problem statement. Section V includes the queuing model of the slices orchestration system. Section VI describes the proposed dynamic resource provisioning solution. Section VII includes simulation results to evaluate the performance of our dimensioning algorithm and to validate the proper operation of the DASA. Finally, Section VIII draws the main conclusions.

## II. Related Works

This section mainly consists of two parts. First, we introduce existing works on network slicing orchestration system. Then, we give a brief background on existing solutions in the literature that leverage the queuing theory in enabling auto scaling systems.

### A. Slices Orchestration Systems

There are several ongoing projects directed towards enabling different aspects of the 5G technology, a number of them are focused on network slicing and how it could be leveraged to realize the 5G technology. For example, while 5G-NORMA is targeted towards enabling the next generation radio access network to support the different use case requirements of the 5G network [13] in a network slicing-centric way, 5GEx also known as 5G Exchange on the other hand is intended towards enabling cross-domain orchestration of network services over multiple administrative domains [14]. Similarly, 5G!Pagoda takes the objectives of 5GEx a step further by presenting a 5G architecture which aligns the vision of both Europe and Japan within the context of the next generation network through the introduction of federated network slicing, which considers the orchestration and provisioning of network slices across multiple continents [16]. Each of the mentioned works presents variant system components based on their fundamental system architectures and functional objectives for the orchestration of network slices.

Similarly, variant algorithms [33] and approaches towards enabling 5G networks based on network slices with a focus on different aspects of the network slice composition have been proposed. While a number of them have focused on the dynamic allocation of the virtualized network resources to network slices and their resource embedding [34], [37]–[39], others have proposed and evaluated their frameworks based on the capacity support for network slices from the network nodes, links and routes that support network slices. For example, the work in [17] takes a look at network slicing from the point of view of network services comprising network Service Function Chains (SFCs) enabled from underlying network infrastructure with network resource constraints in terms of link and node capacity without taking other aspects into account. Taking the idea in [17] a step forward, the work in [20] considers network slicing from a cloud-native approach where network slices are orchestrated from granular modules of legacy network functions to form network of capabilities and services. Similarly, in [19], the authors focus on the orchestration of network slices leveraging a hierarchical architecture consisting of multiple orchestrators using the software defined transport networking scheme.

The work in [13] though with a bit more focus on the realization options of network slicing from a flexible radio access network, proposes yet another system architecture considering also an E2E network slicing perspective for the 5G system orchestration. The project in [2] gives an overarching details on the subject of network slicing while presenting a comprehensive architecture of an E2E network slicing framework leveraging both the concepts of NFV and SDN. It also considers network slicing from the transport network point of view for the backbone network utilizing specifically the concept of SDN. All the presented works have collectively considered different aspects of network slicing from the point of view of orchestrating and connecting virtual network functions (VNFs) in an E2E architectural approach, by considering slice requirements across the entire mobile network segments. However, none of them have presented a system simulation of their orchestration system and how effective they handle network slice orchestration requests under increased workload. In particular, we have not come across any work on multi-domain network slicing that presents an auto-scaling mechanism that manages the autonomous scaling of the orchestration system.

### B. Queuing based Dynamic Auto Scaling Algorithms

The DRP problem has been previously addressed in the context of multi-tier Internet applications [25]. The already proposed solutions are broadly classified into rule-based and model-based approaches. The rule-based approaches are based on reinforcement learning, statistical machine learning, and fuzzy control. On the other hand, the model-based approaches are based on control theory and queuing theory. In contrast to rule-based approaches, model-based approaches require more domain knowledge, but can provide QoS guarantees, while ensuring the system stability [25]. Here we will focus on existing model-based solutions that leverage the queuing theory in enabling auto scaling systems.

There are several works that have tackled the DRP and resource dimensioning problem in the context of the vEPC [26], [27], [29]–[32]. In [31], the authors propose a DRP algorithm for the vEPC considering the capacity of legacy network equipment already deployed. To evaluate the performance of their solution, they model each vEPC element as a M/M/m/K queue and assume that the VNF instantiation time is exponentially distributed. In [26], [30], the authors analyze the performance of a virtualized MME (vMME) with a three-tier design by using a Jacksons network, i.e., a network of M/M/m queues. In those works, the authors use exhaustive search approach to perform the sizing of the number of vMME worker instances. In [32], the authors develop a bi-class (e.g., machine-to-machine -M2M- and mobile broadband -MBB- communications) queuing model for the vEPC. The control plane (CP) and data plane (DP) of the vEPC are respectively modeled as M/M/m/m and M/D/1 nodes. This model constitutes the core of the vEPC-ORA method aimed at optimizing resource assignment for the CP and DP of the vEPC. The authors in [27], [29] propose a heuristic algorithm to carry out the joint resource dimensioning of the vEPC Control Plane (CP) entities. As in this work, that heuristic method relies on a performance model to predict the vEPC response time for a given setup. The authors validate their solution for the planning [27] and DRP [29] of the vEPC.

There exist other works that have also tackled the resource dimensioning problem for particular scenarios such as the Cloud Radio Access Network [49] and the content distribution networks [36]. However, all those aforementioned works are not scenario-agnostic, thus they cannot be applied in the DRP of the NSOS context. Covering this gap, the DRP solutions presented in [10], [28], [35], [44] can be used in a wider range of scenarios.

The authors in [10] propose a DRP solution for multi-tier Internet applications. As in this work, the authors employ a combination of predictive and reactive methods that respectively determine when to provision the resources at large and small time scales. To estimate the performance of the system, each tier instance is modeled as an isolated G/G/1 queue. For each tier, a target response time is set manually and its sizing is carried out regardless of the rest of tiers. In [28], the authors formulate and propose a heuristic to solve the joint optimization problem for the Service Function Chain (SFC) routing and VNF instance dimensioning. In that work, the main objective considered is to minimize the request rejection probability. Similarly, the work in [35] formulates the resource dimensioning problem to minimize the expected waiting time of service chains and proposes a heuristic method to solve it. The authors employ a mixed multi-class BCMP network to model a service chain and estimate its performance metrics by using Mean Value Analysis (MVA) algorithm. Finally, the authors in [44] propose a proactive solution for the dynamic provisioning and flow rerouting of the network services. In contrast to this work, that solution guarantees only the throughput, but not a target mean response time.

The NSOS proposed in this work includes a considerable number of system components (refer to Section III-B). Then, contrary to [10], it is highly desirable that the DRP algorithm driving its auto-scaling performs the resource dimensioning of the different components jointly to guarantee a target response time. In other words, given an overall delay budget, the DRP algorithm should be able to automatically and optimally distribute the overall delay budget among the system components in order to ensure an efficient resources allocation and increase the likelihood of finding a feasible solution. In addition, the execution time of the DRP algorithm should be keep as minimum as possible to maximize the SORs acceptance ratio. In this regard, DRP solutions that rely on MVA algorithm to estimate the performance of the system such as [35] might exhibit high computational complexity. That is because the time complexity of the MVA is $O(N.K)$ [23], where, in the NSOS context, $N$ is the average number of circulating SORs in the network and $K$ is the total number of NSOS entity instances. On the contrary, our DRP solution enables the use of more lightweight performance evaluation techniques like QNA which has time complexity $O(K)$ [23].

## III. An Overview of the proposed NSOS architecture

### A. Main Objectives

One of the main purposes of this work is to develop an efficient global E2E mobile network slicing (including the access, core and transport network parts) orchestration system architecture that takes into account the orchestration of network slices from federated resources. As presented in Fig. 1, the orchestration system shall run seamlessly on virtualized resources for easy adaptability and elasticity. In addition, We also present a queuing model of the interactions between the different building components of the NSOS architecture. Particularly, we would like to build a flexible and scalable orchestration system that strictly respects the service level agreement (SLA) of network slices and most importantly can elastically cope with a given slice orchestration time and service requirements. For this reason, we then further propose an auto scaling algorithm that enables the dynamic resource provisioning for the system to scale autonomously.

When serving slice orchestration requests based on a given Slice Information Graph (SIG) *a.k.a slice blueprint/template*, which defines the slice requirements such as: 1) the type of virtualization technology, 2) the size of the requested E2E mobile network slice, 3) the amount of system resources available for processing the slice requests, 4) the VNF types from which the slice should be orchestrated 5) the slice orchestration domain (for those that should be orchestrated across multiple domains [16], [42]) etc, the system shall scale dynamically in order to provision the slice request within a specified time duration. This implies that when a slice orchestration request reaches the system orchestrator at a given point in time while initially running on certain system resources, depending on the information provided in the SIG, the orchestration system shall dynamically expand if needed in a way that the slice requests will be served efficiently and within a specified processing time. On completion of the request and after some amount of time have elapsed based on the system design, the dynamically added system orchestration resources shall then be released such that only the needed amount of system resources will be running at any given point in time.

Basically, our model is centered around the NSOS components and all the resources needed to operate them efficiently and not on the resources utilized in orchestrating the E2E mobile network slices, thereby making our model a system-centric one. For the sake of simplicity, we would assume that such a global orchestration system would have all of the system resources needed to support the aforementioned slice orchestration requirements as defined in the SIG. Therefore, with this assumption, we would focus on the main system-imposed overhead, which is the slice orchestration latency. Another assumption is that, even though our orchestration framework has all the system resources needed to support any slice requests, for the sake of efficient system resource utilization, they are not always running. Thus, they are instantiated only when needed to ameliorate the slice processing procedure and reduce the orchestration latency in order to ensure the SLA of the requested slice.

### B. Definition of the System Components

As shown in Fig. 1, in our global NSOS architecture, we have considered and introduced a total of nine system entities that will seamlessly operate in a hierarchical order as the fundamental building blocks. Names are carefully selected for each building blocks of the system in order to reflect the parts of the mobile network they are directly handling. Below, we present a high-level description of each of the building blocks.

1) Global Orchestrator (GO): is the component which is responsible for receiving network slice orchestration requests from slice providers and orchestrating the slices in the specified cloud domain in the case of a single domain slice or multiple domains in the case of a federated slice using the optimal Domain-Specific Orchestrator (DSO) or (DSOs), respectively.
2) System Awareness Engine (SAE): is the system component which is responsible for keeping the state, context and the running resources of the entire global orchestration system's entities. It also functions as a global system monitor, which helps keep track of the system's performance.
3) Resource Awareness Engine (RAE): is a system entity which keeps the record of the total resources available on the underlying infrastructure of the orchestration system as well as their respective locations.
4) Domain-Specific Orchestrators (DSOs): is the system component which is basically responsible for orchestrating network slices from a particular administrative domain whose operation spans across a particular network region. Every DSO operating in a particular region has at least one of the following system components needed to actualize a complete E2E network slice orchestration.
5) Domain-Specific Network Function Virtualization Orchestrator (DSNFVO) e.g., the Open Source MANO (OSM): is responsible for communicating directly with the region's DSVIM e.g., Openstack which is in charge of providing virtual resources for the instantiation of virtual network functions.

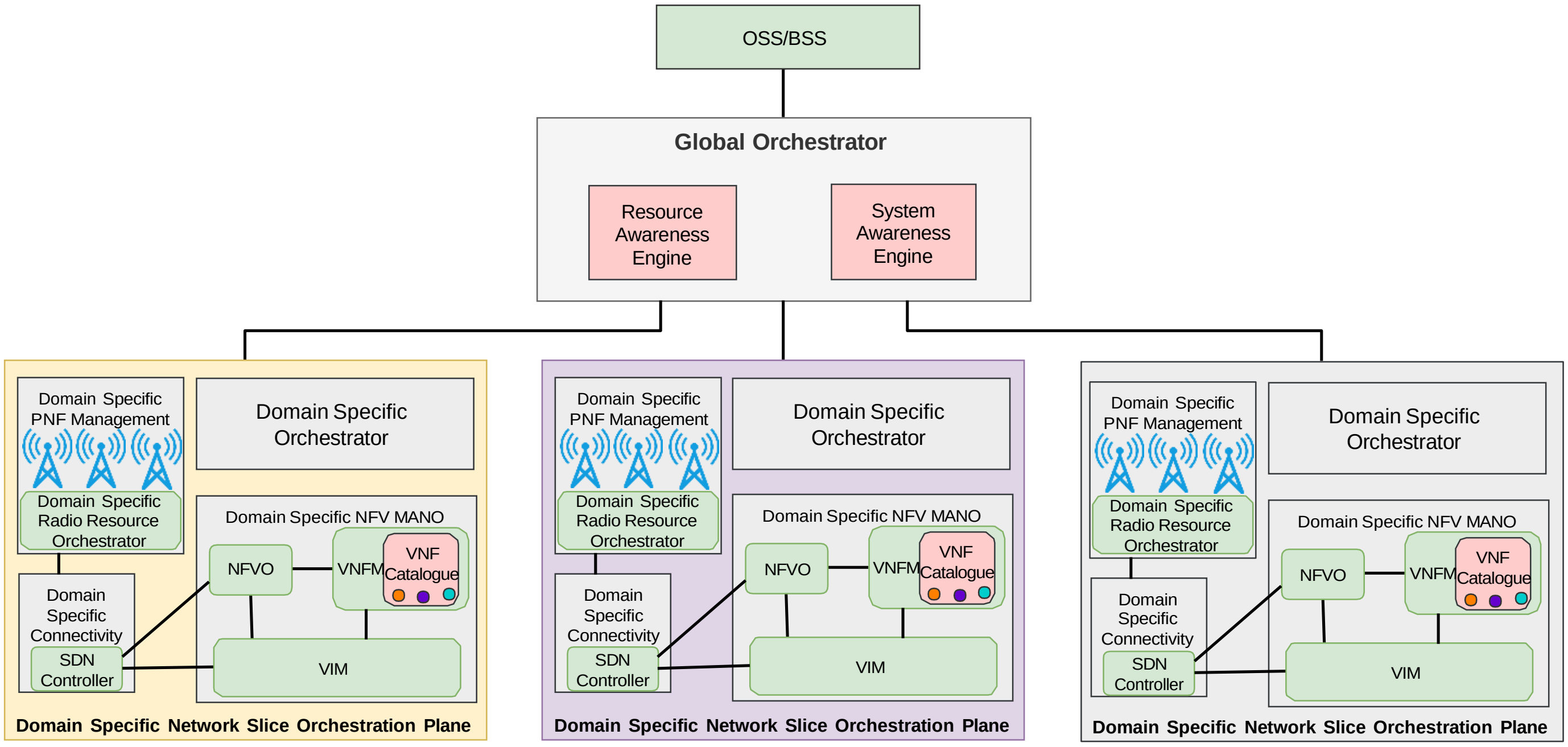

Fig. 1: Simplified practical federated network slice orchestration system deployment architecture.

6) Domain-Specific Radio Resource Orchestrator (DSRRO): is the system component which is solely responsible for orchestrating and allocating radio resources available on already deployed eNBs for the utilization of network slices e.g., the FlexRAN [45].
7) Domain-Specific Software-Defined Networking Controller (DSSDN-C): is the entity responsible for connecting the various orchestrated sub-slices making up a network slice including that of the radio access network [46].
8) Domain-Specific Virtualized Infrastructure Manager (DSVIM) e.g., Openstack: is the system component responsible for the provisioning of virtualized resources to the orchestrated network functions that make up the different sub-slices of a complete mobile network slice.
9) Domain-Specific eNBs (DSeNBs): are the already deployed set of eNBs running in a particular region administered by a DSRRO, running under a particular DSO.

To determine the average rate at which a global slice orchestration system consisting of multiple domain-specific orchestrators would orchestrate federated E2E mobile network slices, we model the system as a queuing system. In the system, there is a GO under whose control is a number of DSO instances and two other system's components dubbed the SAE and RAE. The DSO(s) in turn utilizes the functionalities of three other major components to carry out orchestration tasks. They are the DSNFVO, DSRRO and DSSDN-C as depicted in Fig. 3. The federated slice orchestration procedure follows the activity and sequence diagram shown in Fig. 2 from the perspective of the global orchestrator and from the point of view of each DSO.

Fig. 3 represents a high-level abstraction of the overall orchestration system modeled as a queuing system. In this system, the main point of entry for any slice orchestration request is through the GO. The GO receives the slice request from its North Bound Interface (NBI) as an SIG consisting of all the slice requirements,which the system can support. So, developing our theoretical model around this assumption will help in determining the average time a slice request can be served. In this way, we will be able to develop a system policy that will enable the NSOS to proactively determine the total number of additional instances of the system components needed to orchestrate a slice of a particular set of requirements within a set time duration.

As a result, we see the problem as a job queuing one, in which jobs (federated network slice requests) of certain requirements (therein treated as a black box) arrives at the orchestration system and leaves the system after certain service time had elapsed. However, this job requires the service of multiple servers mostly running in parallel at different stages of the orchestration procedure. Knowing the average service time each part of the job spends in each server (orchestration component) is of utmost importance to us.

### C. *Interactions Between the System Components*

As presented in Fig. 3, when a slice orchestration request arrives at the GO, the request is evaluated based on its requirements and against the running instances of the system components. in order to match the slice's requirements against the running system components instances, the GO consults the SAE through the link A and the SAE replies through the link AR. Based on the reply from the SAE, the GO then consults the RAE on link B to request for the system-wide available resources, in order to determine the available system resources and their locations. Based on the replies from both the SAE and RAE, the request is then forwarded to the concerned DSO(s) via link C. Each DSO then processes its own quota of the federated slice request by further splitting the received SIG into three parts and forwarding each part to the system component responsible for processing each. For example, the part concerning the DSNFVO is sent via link D and similarly through links E and F, both the DSSDN-C and DSRRO receive their quota of the request. Finally, the DSVIM and DSeNBs will process the slice request and reply their DSNFVO and DSRRO via link GR and HR respectively, and forward the reply to their DSO(s) on link CR after which it will be aggregated and forwarded to the GO via link BR. This procedure describes the entire sojourn of processing a federated network slice request.

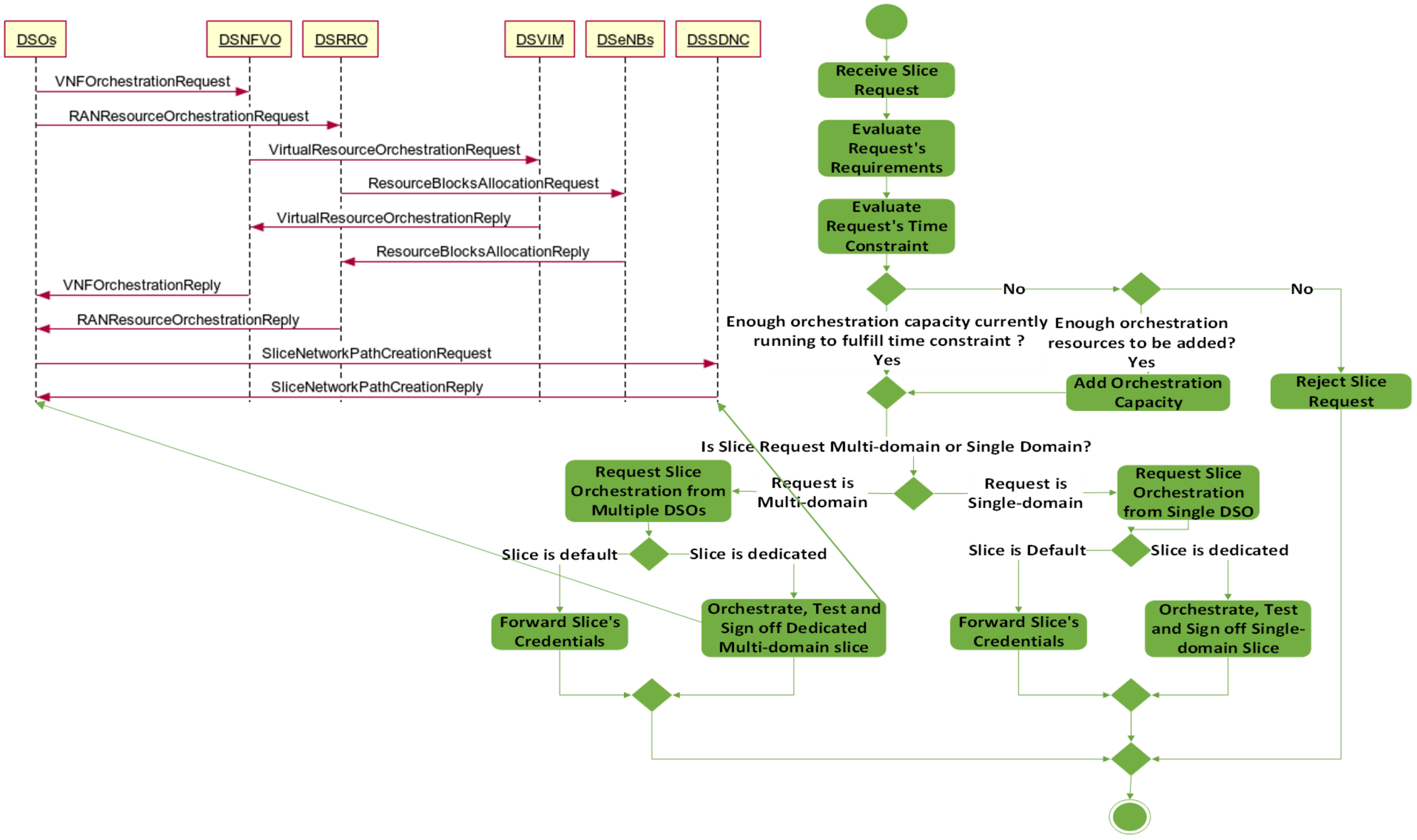

Fig. 2: Activity diagram of the Global Orchestrator.

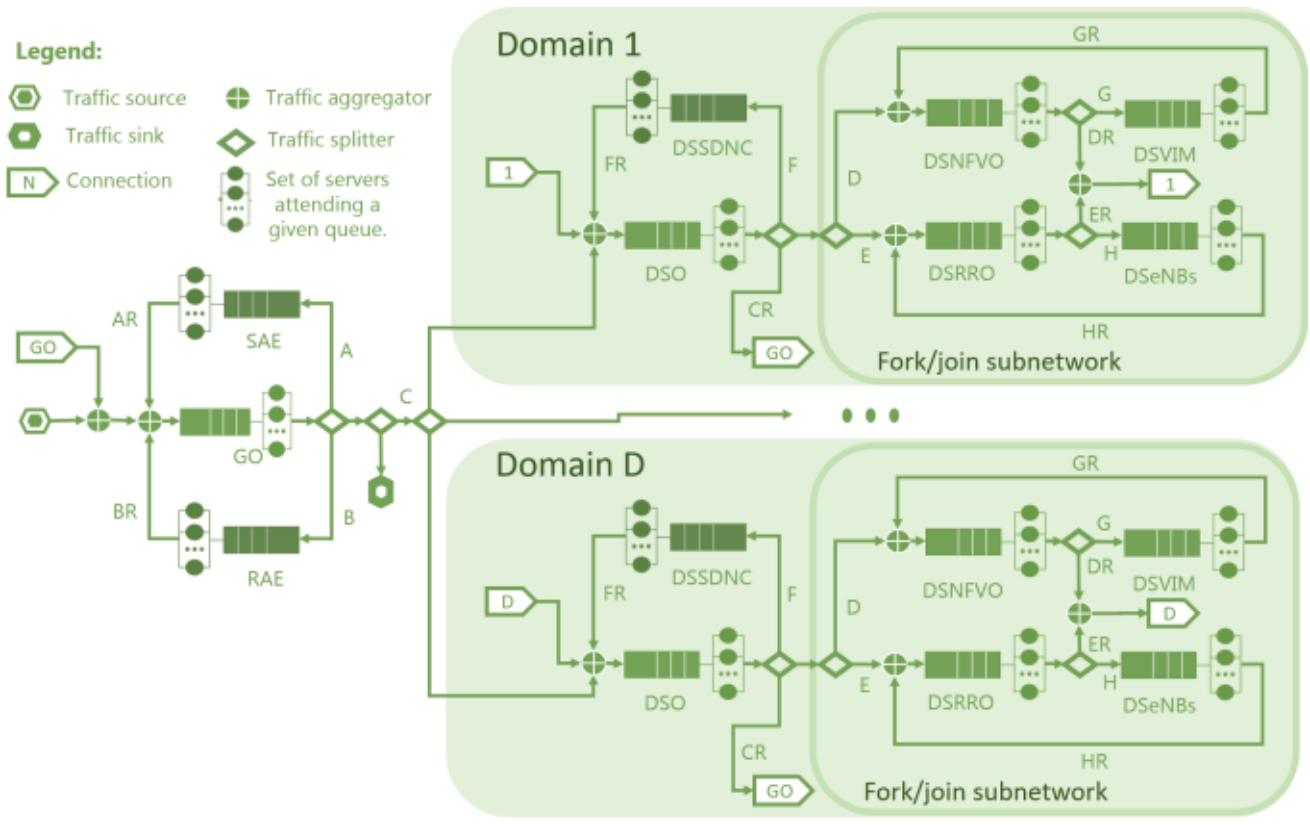

Fig. 3: Queuing model of the Network Slicing Orchestration System.

In line with the architecture, an analytical model of the system is made and results derived from the analysis is fed into our system simulation model in order to determine the adequate number of additional system components that will be instantiated in order to process an E2E network slice within a given duration of time while taking into account the different combinations of slice requirement's SIG compositions. Deducing such figures accurately will be of enormous importance in designing a real efficient E2E federated network slicing orchestration platform that can dynamically scale in order to cope with the slice orchestration requests. In addition to scaling the orchestration resources, this work also takes a progressive approach by not only presenting a comprehensive understanding of the concept of network slicing from an E2E orchestration perspective but also incorporates the idea of the resource affinity paradigm as discussed in[18] in order to develop a system model for an elastic orchestration platform and also include other interesting aspects.

## IV. Problem Statement

In this section we will describe our system model and problem statement. From a higher abstraction point of view, the NSOS detailed in the previous section can be regarded as a set of entities $E = \{GO, SAE, RAE, DSO_d, DSNFVO_d, DSVIM_d, DSSDNC_d, DSRRO_d, DSeNBs_d\}$ interacting among them. Please note that the subindex $d \in [1, 2, .., D]$ is included to specify the domain, which may be associated with a specific geographical region, the entity belongs to. There might be several instances per entity $e \in E$. Each entity instance is deployed on an individual Virtual Machine (VM). To serve the different slice orchestration requests (SORs), these entities interact with each other by exchanging signaling messages. In general, an instance $i$ of a given entity $e \in E$ will execute a task, which consume computational resources (e.g., CPU time, RAM, disk, network bandwidth), to process every incoming packet. Here, we assume that the CPU is the resource acting as bottleneck of the different entities. Once the packet has been processed, the entity instance generates and sends the corresponding reply message according to the specific SOR call flow.

DRP refers broadly to enabling the automation of the resources scaling of a given system depending on the current or foreseen workload in the near future so that a set of performance requirements are met. One of the main objectives of this work is to propose a DRP solution for our NSOS. Specifically, the aim is to provide a solution that allows the NSOS to autonomously adapt its own computational and virtual resources to handle workload fluctuations, while keeping the desired performance to serve the SORs. Here, we will consider as performance requirement that the mean response time of the NSOS $T$ to serve a SOR has to be kept under a threshold $T_{max}$.

One of the main components of a DRP solution is the resources dimensioning module, which is in charge to determine how much to provision for a given workload [10]. The resource dimensioning problem for the NSOS might be formulated as follows:

$$minimize \quad F(\boldsymbol{m}) = \left( \sum_{e \in E} \sum_{i} m_i^{(e)} \right) \tag{1a}$$

**Constraints** :

$$C1: \quad T(\boldsymbol{m}) \leq T_{max}, \tag{1b}$$

$$C2: \quad m_i^{(e)} \leq m_{max}^{(e)} \quad \forall \quad k \in [1, K] \cap \mathbb{N} \tag{1c}$$

The decision variables of the optimization problem are the number of physical CPU cores $m_i^{(e)}$ allocated to each instance $i$ of a given entity $e \in E$. Objective (1a) aims to minimize the amount of computational resources (expressed in number of CPU cores) to be allocated to the NSOS. This objective is relevant because it is equivalent to minimize the energy consumption and operational costs of the NSOS. Constraint (1b) guarantees that the actual mean response time of the NSOS to serve the slice orchestration requests is below a mean delay threshold $T_{max}$. Constraint (1c) limits the maximum number of physical cores allocated to a single entity instance. To have a single instance would be optimal for minimizing the amount of required resources (statistical multiplexing). However, each physical machine has a maximum number of physical cores and they are shared among several VMs. Constraint (1c) facilitates the bin packing problem in the context of network embedding.

Noteworthy, the resource dimensioning problem formulated above can be solved by means of a exhaustive method search as follows. The brute force algorithm could determine the initial processing instances allocation to each entity using the condition for the system stability, i.e., $\boldsymbol{m_e} = \lceil \boldsymbol{\lambda_e} / \boldsymbol{\mu_e} \rceil$, where $\boldsymbol{m_e}$, $\boldsymbol{\lambda_e}$, and $\boldsymbol{\mu_e}$ are vectors containing respectively the number of processing instances allocated, the aggregated arrival rate, and the packet processing rate per processing instance for each entity. Then, assuming there are $N_e$ different entities in the system, $M_0 = \sum_{i=1}^{N_e} \boldsymbol{m_e}(i)$ processing instances are allocated in the initial assignment to fulfill the stability condition. Then the brute force algorithm iterates until the mean response time of the NSOS is below the maximum delay threshold, i.e., $T < T_{max}$. At each iteration it increments by one the total number of processing instances allocated to the whole system, $M$. It will then check every combination to allocate the $M - M_0$ processing instances among the entities of the system. After performing the check, It will choose the allocation that achieves the minimum mean response time of the system. Despite the simplicity of the exhaustive search method, it requires

$$N_{checks} = \sum_{M=M_0}^{M^*} \binom{N_e}{M - M_0} = \sum_{M=M_0}^{M^*} \frac{(N_e + M - M_0 - 1)!}{(M - M_0)!(N_e - 1)!}$$

evaluations of the NSOS mean response time, which is impractical.

## V. NSOS'S MEAN RESPONSE TIME ESTIMATION USING QUEUING THEORY

The proposed model to estimate the performance of our slices orchestration system consists of an open network of G/G/m queues, where each queue represents an instance of a given entity of the system like the GO or a DSO (see Fig. 3). Each instance of a given entity runs on a separated virtualization container or virtual machine (VM). For the sake of simplicity, only one instance per entity is depicted in Fig. 3. In Kendall's notation, a G/G/m queue is a queuing node with $m$ servers, arbitrary arrival and service processes, FCFS (First-Come, First-Served) discipline, and infinite capacity and calling population. The system is modeled by imagining a 3GPP-like service-based architecture where each components of the 5G core is considered an instantiated service and more instances of the same service can be launched as at when needed based on the system load.

This modeling approach has been previously proposed and validated for virtualized network entities in [23]. Please observe that the proposed model is intended to carry out the computational resources dimensioning of the system. A model targeted to accurately assess the performance of the system should take into account additional elements such as the communication links between the system entities.

As stated earlier, each queue stands for an instance $i$ of the entity of the set $E \in \{GO, SAE, RAE,$

TABLE I: Model parameters.

| Notation | Description |
|---|---|
| $K$ | number of G/G/m queues |
| $P$ | steady state transition probability matrix |
| $k, i$ | network nodes indexes |
| $p_{ki}$ | probability of a packet leaving a node $k$ to node $i$ |
| $p_{0k}$ | probability that a packet leaves the network |
| $\lambda_{0k}$ | mean arrival rate of the external arrival process at queue k |
| $c^2_{0k}$ | squared coefficient of variation of the external arrival process at queue k |
| $\mu_k$ | mean service rate of each server at queue k |
| $c^2_{sk}$ | squared coefficient of variation of service processes at queue k |
| $c^2_{ak}$ | squared coefficient of variation of the aggregated arrival process at queue k |
| $a_k$, $b_{ik}$ | Coefficients of the set of linear equations to estimate the SCVs of the aggregated arrival process at each queue $k$. |
| $\omega_k$, $x_i$, $\gamma_k$ | Auxiliary variables when $a_k$ and $b_{ik}$ are computed. |
| $q_{0k}$ | the proportion of arrivals to node k from its external arrival process |
| $q_{ik}$ | the proportion of arrivals to node k from node i |
| $m_i$ | number of servers of node i |
| $c^2_{si}$ | squared coefficient of variation of service processes at queue i |
| $\rho_k$ | the utilization of the node $k$ |
| $T_k$ | Mean system response time of node $k$ |
| $W_k$ | Mean waiting time of node $k$ |
| $\beta$ | The Kraemer and Langebach-Belz approximation |
| $W_k^{M/M/m}$ | the mean waiting time for an M/M/m queue |
| $C(m, \rho)$ | the Erlang's C formula |
| $T$ | the overall mean response time |
| $V_k$ | the average number of visits to node $Q_k$ |

$DSO_d$, $DSNFVO_d$, $DSVIM_d$, $DSSDNC_d$, $DSRRO_d$, $DSeNBs_d$ }, where the subindex $d \in [1, 2, .., D]$ is included to specify the domain, which may be associated with a specific geographical region, the entity belongs to. Please note that the three solid circles in Fig. 3 stand for domains with indexes from 2 to $D-1$. As described in Section III, the GO orchestrates the slices of different domains. Each domain has its own dedicated DSO, DSNFVO, DSVIM, DSSDNC, DSRRO, and DSeNBs entities, which are in charge of orchestrating and managing the slices of a given geographical region. In our model, entities belonging to different domains are treated separately, though they may have the same functionality.

A queue may have $m_i^{(e)}$ servers that represent processing instances allocated to the corresponding instance $i$ of the entity $e \in E$. These processing instances (i.e., virtual CPUs running on physical CPU cores) process messages from the same queue. Please note that each entity has to process one or several messages to serve every incoming slice orchestration request as shown in Fig. 2. We assume there is a sentry per entity that distributes the workload (messages) among the entity instances according to their processing capacity.

## A. System Response Time Estimation

The performance metric considered in this work that drives the auto-scaling process of the slices orchestration system is the mean response time. That is, when possible, the system will scale up or down its computational resources depending on the foreseen workloads peaks in the near future so that a given mean response time to serve the slice orchestration requests is guaranteed. Consequently, based on the aforementioned proposed queuing model, we need to derive the mean response time of the system.

To that end, we use the approximated technique proposed in [22] for the Queuing Network Analyzer, hereinafter referred to as QNA method. This methodology was applied and validated to estimate the mean response time of a VNF with several components in [23] proving it outperforms the standard queuing techniques of analysis in terms of estimation error.

It shall be noted that to capture some particularities in the operation of our system we need to include some considerations in the analysis that differs from the originally proposed QNA method. For instance, the DSO entity sends requests in parallel to the DSNFVO and DSRRO entities for the reservation/allocation of resources for a given slice orchestration request (see Fig. 2). This blocks the call flow at the DSO for this slice orchestration request until both the DSNFVO and DSRRO answer the request. This behavior is captured by modeling the subnetwork composed of the DSNFVO, DSVIM, DSRRO, and DSeNBs for a given domain $d$ as a fork/join subnetwork with two parallel branches (e.g., DSNFVO/DSVIM and DSRRO/DSeNBs). Then, the branch with the highest response time will determine the response time of the fork/join subnetwork.

In order to estimate the mean response time in a systematic manner, let $K_e$ denote the number of instances for each entity $e$ of the set $E$. Then, we have $K = \sum_{e \in E} K_e$ instances in the system and, thus, K queues in the queuing model. To simplify the notation in this analysis, we map each entity instance to an integer index $k \in [1, K]$. The specific assignment of index to each entity instance does not affect the subsequent analysis. The following input parameters are required to estimate the mean response time of the system: i) the steady state transition probabilities matrix $P = [p_{ki}]$, where $p_{ki}$ denotes the probability of a packet leaving the node $k$ to the next node $i$ or leaves the network with probability $p_{0k} = 1 - \sum_i p_{ki}$; ii) the mean and squared coefficient of variation (SCV) of the external arrival processes at queue $k$, $\lambda_{0k}$ and $c^2_{0k}$; and iii) the mean and SCV of the service processes at queue $k$, $\mu_k$ and $c^2_{sk}$.

*1) Internal flows parameters estimation:* The mean arrival rate to each queue $k$, $\lambda_k$, can be computed by solving the flow balance equations:

$$\lambda_k = \lambda_{0k} + \sum_{i=1}^{K} \lambda_i \cdot p_{ik} \quad (2)$$

The most interesting aspect of the QNA method is that it estimates the SCV of the aggregated arrival process to each queue $c^2_{ak}$ from a set of linear equations. Then, it is only required to monitor the external arrival processes to the system, which is the arrivals of new slices' orchestration requests to the GO in our case, to estimate the second order moment of the arrival process at each entity instance. Specifically, we can estimate the SCV of the aggregated arrival process to each entity instance solving the following set of linear equations:

$$c^2_{ak} = a_k + \sum_{i=1}^{K} c^2_{ai} b_{ik}, \quad 1 \le k \le K \quad (3)$$

$$a_k = 1 + \omega_k \left( q_{0k} c^2_{0k} - 1 \right) + \sum_{i=1} q_{ik}[(1 - p_{ik}) + p_{ik}\rho^2_i x_i] \quad (4)$$

$$b_{ik} = \omega_k q_{ik} p_{ik} (1 - \rho^2_i) \quad (5)$$

$$x_i = 1 + m_i^{-0.5}(max\{c^2_{si}, 0.2\} - 1) \quad (6)$$

$$\omega_k = \left(1 + 4(1 - \rho_k)^2(\gamma_k - 1)\right)^{-1} \quad (7)$$

$$\gamma_k = \left(\sum_{i=0}^{K} q^2_{ik}\right)^{-1} \quad (8)$$

To simplify the computation of the $c^2_{ak}$, the QNA method employs approximations. Specifically, it uses a convex combination of the asymptotic value of the SCV $(c^2_{ak})_A$ and the SCV of an exponential distribution ($c^2_{exp} = 1$), i.e., $c^2_{ak} = \alpha_k(c^2_{ak})_A + (1 - \alpha_k)$. The asymptotic value can be found as $(c^2_{ak})_A = \sum_{i=1}^{K} q_{ik} c^2_{ik}$, where $q_{ik}$ is the proportion of arrivals to $Q_k$ that came from $Q_i$. That is, $q_{ik} = \frac{\lambda_i v_i p_{ik}}{\lambda_k}$. And $\alpha_k$ is a function of the queuing node utilization $\rho_k = \frac{\lambda_k}{\mu_k m_k}$ and the arrival rates. This approximation yields the above set of linear equations, which may be solved to get $c^2_{ak}$, $\forall \{k \in \mathbb{N} | 1 \leq k \leq K\}$. Last, in the above equations $q_{0k} = \lambda_{0k}/\lambda_k$.

*2) Mean response time computation per node:* Once the $\lambda_k$ and $c^2_{ak}$ for the aggregated arrival process to each node $k$ are estimated, we can compute the mean queuing waiting time for each node $k$, $W_k$.

If the node $k$ has only one server (or one processing instance allocated, $m_i^{(e)} = m_k = 1$), $W_k$ can be estimated as:

$$W_k = \frac{\rho_k \cdot (c^2_{ak} + c^2_{sk}) \cdot \beta}{2 \cdot \mu_k(1 - \rho_k)} \quad (9)$$

with

$$\beta = \begin{cases} exp\left(-\frac{2\cdot(1-\rho_k)\cdot(1-c^2_{ak})^2}{3\cdot\rho_k\cdot(c^2_{ak}+c^2_{sk})}\right) & c^2_{ak} < 1 \\ \beta = 1 & c^2_{ak} \geq 1 \end{cases} \quad (10)$$

If, by contrast, the node $k$ is a GI/G/m queue ($m_i^{(e)} = m_k = m$), $W_k$ can be estimated as:

$$W_k = 0.5 \cdot \left(c^2_{ai} + c^2_{si}\right) \cdot W_k^{M/M/m} \quad (11)$$

where $W_k^{M/M/m}$ is the mean waiting time for a M/M/m queue, which can be computed as:

$$W_k^{M/M/m} = \frac{C(m_k, \frac{\lambda_k}{\mu_k})}{m_k\mu_k - \lambda_k} \quad (12)$$

and $C(m, \rho)$ represents the Erlang's C formula which has the following expression:

$$C(m, \rho) = \frac{\frac{(m\cdot\rho)^m}{m!} \cdot \frac{1}{1-\rho}}{\sum_{k=0}^{m-1} \frac{(m\cdot\rho)^k}{k!} + \frac{(m\cdot\rho)^m}{m!} \cdot \frac{1}{1-\rho}} \quad (13)$$

*3) Global Response Time Computation:* Let $V_k$ denote the visit ratio for the node $k$ ($Q_k$) which is defined as the average number of visits to node $Q_k$ by a packet during its lifetime in the network. That is, $V_k = \lambda_k/(\sum_{k=1}^{K} \lambda_{0k})$. And let $\mathrm{K}_e \subset \{1, ..., K\}$ with $|\mathrm{K}_e| = K_e$ be the subset of indexes associated with the instances of the entity $e$. Then, we can compute the mean response time for each entity, $T_e$, as:

$$T_e = \sum_{k \in \mathrm{K}_e} (W_k + \frac{1}{\mu_k}) \cdot V_k \quad (14)$$

As stated earlier, the $DSNFV\,O_d$, $DSV\,IM_d$, $DSRRO_d$, and $DSeNBs_d$ entities of the domain $d$ are modeled as a fork/join subnetwork of queues with two parallel branches. The mean response time of the fork/join subnetwork of the domain $d$, $T_d^{(FJS)}$ will be given by:

$$T_d^{(FJS)} = max(T_{DSNFV\,O_d} + T_{DSV\,IM_d}, T_{DSRRO_d} + T_{DSeNBs_d}) \quad (15)$$

Finally, the overall mean response time $T$ can be estimated as:

$$T = T_{GO} + T_{SAE} + T_{RAE} + \sum_{d=1} T_{DSO_d} + T_d^{(FJS)} \quad (16)$$

### *B. Transtion Probabilities for the Slices Orchestration System*

As stated earlier, the steady-state transition probabilities are input parameters to estimate the mean response time of the system. They can be derived directly from the system operation.

Let $V_e$ denote the visit ratio of the entity $e \in E$ which is defined as the average number of visits to entity $e$ by a SOR during its lifetime in the NSOS. That is,

$$V_e = \lambda_e / \sum_e \lambda_{0e} = \lambda_e/(\lambda_{0GO}) \quad (17)$$

The visit ratios and the transition probabilities are related through (2) (flow balance equations):

$$V_e = \frac{\lambda_{0e}}{\sum_{e \in E} \lambda_{0e}} + \sum_{es \in E} V_{es} \cdot p_e^{es} \quad (18)$$

where $p_e^{es}$ denotes the transition probability from entity $es \in E$ to entity $e \in E$. The visit ratios of the $GO$, $RAE$, and $SAE$ entities are given by the number of messages they have to process for every slice orchestration request. Then, $V_{GO} = 3$, $V_{SAE} = 1$, and $V_{RAE} = 0$ (see Fig. 2). Let $\alpha_d$ denote the percentage of the total incoming slices orchestration requests to the system which are addressed to the domain $d$. Then, the visit ratios of the entities $DSO_d$, $DSSDNC_d$ $DSNFV\,O_d$, $DSV\,IM_d$, $DSRRO_d$, and $DSeNBs_d$ will be $\alpha_d$ times the number of messages these entities have to process for every incoming slice orchestration request, i.e., $V_{DSO_d} = alpha_d \cdot 3$, $V_{DSNFV\,O_d} = alpha_d \cdot 2$, $V_{DSV\,IM_d} = alpha_d$, $V_{DSRRO_d} = alpha_d \cdot 2$, $V_{DSeNBs_d} = alpha_d$, and $V_{DSSDNC} = \alpha_d$.

Additionally, the sum of the transition probabilities for a given entity $e$ are normalized to unity:

$$p_{0e} + \sum_{ed \in E} p^e_{ed} = 1 \quad (19)$$

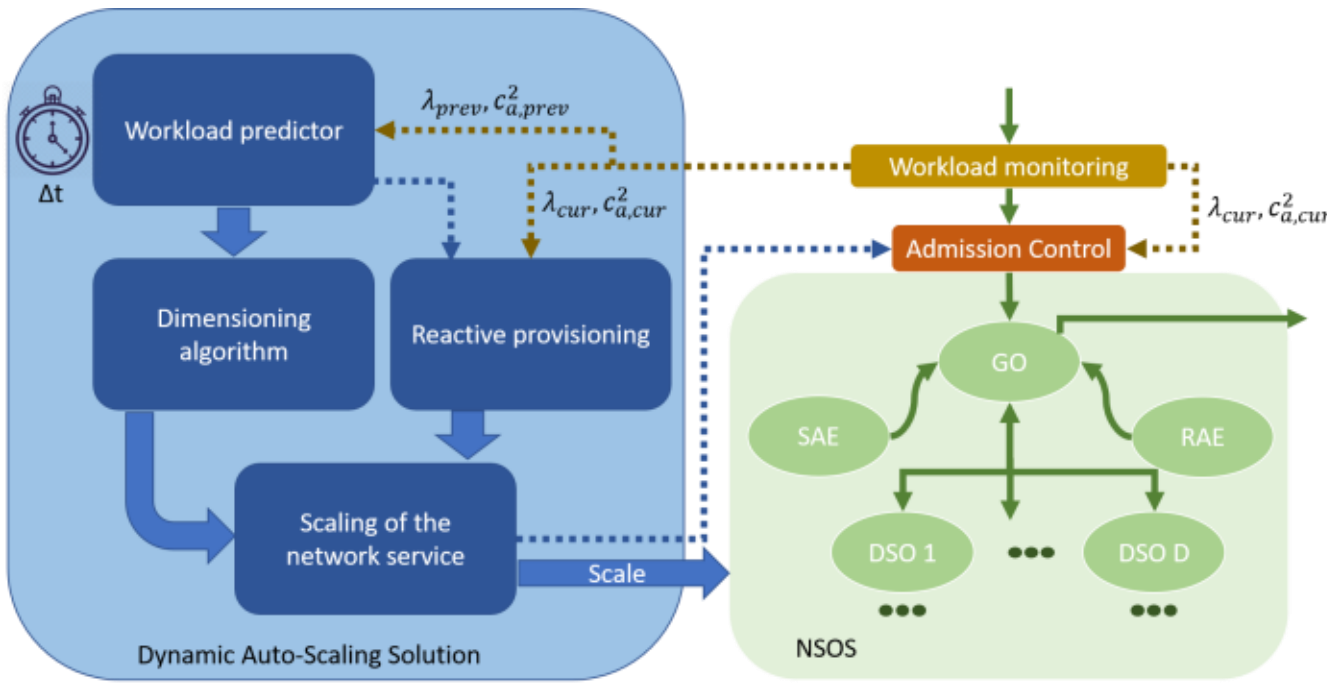

Fig. 4: Dynamic Auto-Scaling Solution for the NSOS.

In our case, we can compute the transition probabilities between entities for the slices orchestration system model using (18) and (19). Specifically, we get

$$p^{GO}_{SAE} = \frac{1}{3}; \quad p^{GO}_{RAE} = 0; \quad p^{GO}_{DSO_d} = \alpha_d \cdot \frac{1}{3}; \tag{20}$$

$$p^{SAE}_{GO} = 1; \quad p^{RAE}_{GO} = 1; \tag{21}$$

$$p^{DSO_d}_{DSNFVO_d} = \frac{1}{3}; \quad p^{DSO_d}_{DSSDNC_d} = \frac{1}{3}; \quad p^{DSO_d}_{GO} = \frac{1}{3}; \tag{22}$$

$$p^{DSNFVO_d}_{DSO_d} = 0.5; \quad p^{DSNFVO_d}_{DSVIM_d} = 0.5; \tag{23}$$

$$p^{DSVIM_d}_{DSNFVO_d} = 1; \quad p^{DSSDNC_d}_{DSO_d} = 1; \tag{24}$$

Finally, assuming that the workload is distributed among the instances of a given entity $e$ according to its processing capacity, the transition probability $p^{e1_i}_{e2_j}$ from the instances $i$ of the entity $e1 \in E$ to the instance $j$ of the entity $e2 \in E$ can be simply computed as

$$p^{e1_i}_{e2_j} = \frac{m^{(e2)}_j}{\sum_l m^{(e2)}_l} \cdot p^{e1}_{e2} \tag{25}$$

## VI. Dynamic Auto-Scaling Solution of NSOS

The section describes a novel Dynamic Resource Provisioning solution for the NSOS. The proposed DRP solution (see Fig. 4) enables the NSOS to adapt its own resources depending on the near future foreseen workload so that its mean response time $T$ is kept under the delay threshold $T_{max}$ (i.e., $T \leq T_{max}$). Figure 4 shows the main blocks of the DRP solution for the NSOS, which are explained in the following subsections.

### A. Workload predictor

This block is responsible for estimating the peak demand for the NSOS until the next decision to provision is taken. The workload predictor is executed synchronously every $\Delta t$ units of time. The value of $\Delta t$ could be established from statistics of the workload arrival process in order to find a balance between the rate of scaling requests issued by the DRP module and resources savings. The workload predictor might be implemented by using Artificial Intelligence (AI) techniques, as used in [42], such as machine learning. This block receives as input the statistics of the peak traffic workload arriving at the NSOS (*e.g.*, mean arrival rate $\lambda_{0GO,prev}$ and SCV of the packet inter-arrival times $c^2_{0GO,prev}$) during the last period $\Delta t$. These statistics are measured by a workload monitoring agent and reported to the DRP module every $\Delta t$ units of time. As output, this block provides the predicted values of the mean arrival rate $\lambda_{0GO}$ and the SCV of the packet inter-arrival times $c^2_{0GO}$ of the peak traffic demand for the next period of length $\Delta t$.

### B. Dimensioning algorithm

This block is in charge of the sizing of the computational, network, and the number of virtualization containers (VM and or OS-level containers) from $\lambda_{0GO}$ and $c^2_{0GO}$ so that $T \leq T_{max}$ during the next period of length $\Delta t$.

To solve the resources dimensioning problem formulated in Section IV, we employ a heuristic method that relies on the analytical model previously described in Section V (see Algorithm 1). As input, the algorithm requires the target mean response time $T_{max}$, the maximum number of available physical CPU cores $M_{max}$, the mean and SCV of the external arrival process provided by the predictor (i.e., $\lambda_{0GO}$ and $c^2_{0GO}$), and the mean and SCV $\boldsymbol{\mu}$, and $\boldsymbol{c^2_s}$. The inputs $\boldsymbol{\mu}$ and $\boldsymbol{c^2_s}$ are column vectors containing respectively the mean service rate and the SCV of the service times for all the entities in the set $E$. The mean service rate $\mu_e$ and the SCV of the service times $c^2_{es}$ per entity can be measured offline for a given processing instance type.

**Algorithm 1** Dimensioning Algorithm

**Input:** $T_{max}$, $M_{max}$, $\lambda_{0GO}$, $c^2_{0GO}$, $\boldsymbol{\mu_e}$, $\boldsymbol{c^2_{es}}$.
**Output:** Required number of instances (or virtualization containers) $\boldsymbol{I}$ per entity and the processing instances to be allocated to each entity $\boldsymbol{m}$.

**Initialization** Compute $\boldsymbol{\lambda_e}$ using (17); $\boldsymbol{m} \leftarrow \lceil \boldsymbol{\lambda_e} \;\; \boldsymbol{\mu_e} \rceil$; $M_0 \leftarrow \sum_{e \in E} \boldsymbol{m}(e)$; $M \leftarrow min\{M_0, M_{max}\}$; $\boldsymbol{K} \leftarrow \lceil \boldsymbol{m} \;\; \boldsymbol{m_{max}} \rceil$; Compose the network of queues and compute $\boldsymbol{P}$ using (20)-(25); Compute internal flows parameters $\boldsymbol{\lambda}$ and $\boldsymbol{c^2_a}$ using (2)-(8); $T$ estimation given the initial stability conditions using (9)-(16);
**while** $T > T_{max}$ or $M < M_{max}$ **do**
  $\boldsymbol{m_{aux}} \leftarrow \boldsymbol{m} + \boldsymbol{1}_{N\times 1}$; $\boldsymbol{K_{aux}} \leftarrow \lceil \boldsymbol{m} \;\; \boldsymbol{m_{max}} \rceil$;
  Recompose the network of queues for $\boldsymbol{m_{aux}}$ and $\boldsymbol{I_{aux}}$; and compute $\boldsymbol{P}$ using (20)-(25).
  Recompute $\boldsymbol{\lambda}$ and $\boldsymbol{c^2_a}$ using (2)-(8);
  Estimate the queuing waiting times vector $\boldsymbol{W_{aux}}$ using (9)-(13) and considering the above input parameters;
  $e^* \leftarrow \underset{k \in [1,K] \cap \mathbb{N}}{\text{argmax}} (\boldsymbol{W}(e) - \boldsymbol{W_{prev}}(e))$;
  $\boldsymbol{m}(e^*) \leftarrow \boldsymbol{m}(e^*) + 1$; $\boldsymbol{K}(e^*) \leftarrow \lceil \boldsymbol{m}(e^*)/m^{(e)}_{max} \rceil$;
  $T \leftarrow T - \boldsymbol{W_{prev}}(e^*) + \boldsymbol{W}(e^*)$; $M \leftarrow M + 1$;
**end while**

The Algorithm 1 searches for the minimum number of processing instances to be allocated to the network service so that $T < T_{max}$. The algorithm iterates until either $T \leq T_{max}$ or $M \geq M_{max}$. At each iteration it seeks for the entity $e^* \in E$ that most contributes in the reduction of $T$ when one additional processing instance is allocated to such entity $e^*$. Then, the algorithm actually assigns one additional processing instance to entity $e^*$.

In contrast to the brute force algorithm described in Section IV, Algorithm 1 requires only $(M^* - M_0)$ evaluations of the performance model to find the solution. Where $M^*$ is the total number of required physical CPU cores estimated by the algorithm, and $M_0$ denotes the initial allocation of CPU cores to fulfill the stability condition. Let $M_{max}$ denote the maximum number of available CPU cores. Considering that QNA method takes linear time $O(K)$ with the size of the network $K$ (i.e., number of queuing nodes) as reported in [23]. Then, the time complexity of Algorithm 1 in the worst-case scenario is $O((M_{max} - M_0) \cdot K^*)$, where in our case $K^*$ is the required number of instances for all the entities, which is estimated by the algorithm. It is difficult to determine how the algorithm complexity depends on its input parameters. That is because the complexity is a function of the algorithm output. To overcome this obstacle, we provide an experimental evaluation of the algorithm time complexity in Section VII-B showing that it is $O\left(\lambda_{0GO}^2 \cdot log\left(T_{max}^{-1}\right)\right)$ when $M^* < M_{max}$. We also provide an analytical derivation of the algorithm time complexity in Appendix **??** for a hypothetical case where $N_e = 1$ which provide useful insights on how it depends on the different input parameters.

Observe that although we have particularized the proposed resources dimensioning heuristic for the performance model developed in Section V, it could be used a different performance model or even a distinct approach to estimate the response time of the NSOS. Accordingly, the input parameters, except $T_{max}$ and $M_{max}$, could change.

Please note that denotes the Hadamard division, which is defined as the element wise division, of two vectors in Algorithm 1. In the same way, the operator $\lceil \cdot \rceil$ refers to the element wise ceiling operation for a vector.

Finally, the output of the dimensioning algorithm is used by the scaling subsystem, which will be in charge of allocate or release the corresponding resources.

### C. Scaling of the network service

This block is in charge of initiating the required procedures for allocating or releasing the network service resources. As input, it uses the required processing instances per entity $\boldsymbol{m}$ provided by either the dimensioning algorithm or the reactive provisioning block whose functionality is described in the next bullet point. It keeps track of the resources currently allocated to the network services. Then, it can determine how much resources have to be reserved or freed given the output of the dimensioning algorithm or the reactive provisioning block.

### D. Reactive provisioning

This block receives frequently the current statistics of the traffic demand ($\lambda_{cur}$ and $c^2_{a,cur}$) measured by the workload monitoring agent. Its mission is to trigger asynchronous resources scaling requests when it detects an unexpected workload surge that has not been foreseen by the workload predictor.

### E. Request policing

It is worthy to note that the DRP module interacts with the admission control procedure of the NSOS. The admission control procedure enables the NSOS to decline excess SORs during temporary overloads. To that end, the admission control procedure might use the current statistics of the traffic demand ($\lambda_{cur}$ and $c^2_{a,cur}$) provided by the workload monitoring agent and the information of current capacity of the network service to serve requests. Although the reactive provisioning block will react before unexpected workload surges, the reaction time might be non-negligible (execution time of the reactive provisioning algorithm, time to carry out the procedures of resources reservation, execution time of the resources embedding algorithm, time to instantiate new VM, etc.). Then, it is required an admission control mechanism to guarantee that the performance requirements for the network service are met all the time.

## VII. Results

In this section, we assess the performance of our DRP solution based on simulations. First, we present an evaluation of our resources dimensioning algorithm. Second, we show the proper operation of our DRP solution in a typical scenario.

### A. Experimental setup

To validate the proposed queuing model and operation of our DRP algorithm solution, we employed a queuing simulator of the slices orchestration system, which was developed in the Matlab Simulink environment. Table II includes the configuration of the main parameters for all the simulations carried out in the subsequent experimental evaluation. Except where otherwise noted, we assume that the workload is equally distributed among the different DSOs.

This simulator implements the call flows triggered by each slice orchestration request (see Fig. 2). Each entity instance is simulated as a FCFS queue with one or several generic servers. The specific number of servers for any entity instance depends on the processing instances allocated to it for the given simulation. For all the processing instances, no matter which entity they were allocated to, we assumed the same service time distribution. The shape of this distribution was derived from real measurements of the service times for a VNF that have a similar operation to the entities of the slices orchestration system. The service time distribution for each processing instance has a mean of 100 $\mu s$ (or, equivalently, a service rate equal to 10000 $packets/second$) and a SCV equal to 0.65.

To the best of the authors' knowledge, there is no work in the literature addressing the characterization of the slice orchestration requests generation process. Then, in our simulations, we supposed that the slice orchestration requests arrive at the GO entity according to a Poisson distribution. The aggregated arrival rate was modulated according to the temporal distribution measured in [24] for the aggregated mobile traffic. Specifically, the temporal distribution of the workload profile considered is depicted in Fig. 5.

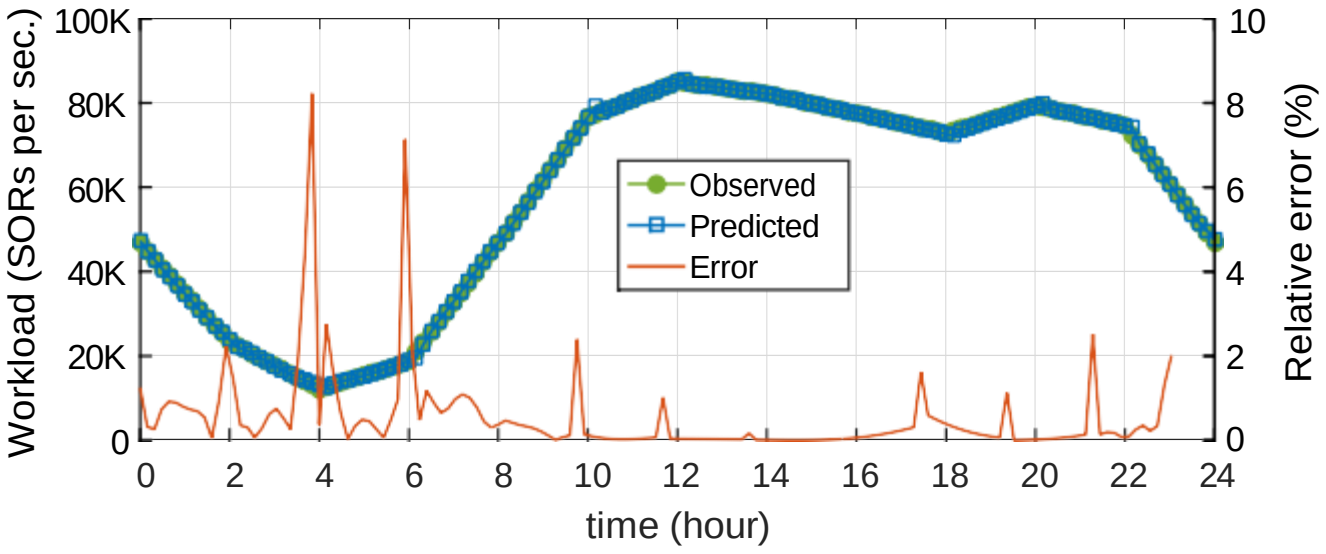

Fig. 5: The temporal distribution of the workload profile (expressed in slices orchestration resquests -SORs- per second) and the prediction error of the workload predictor.

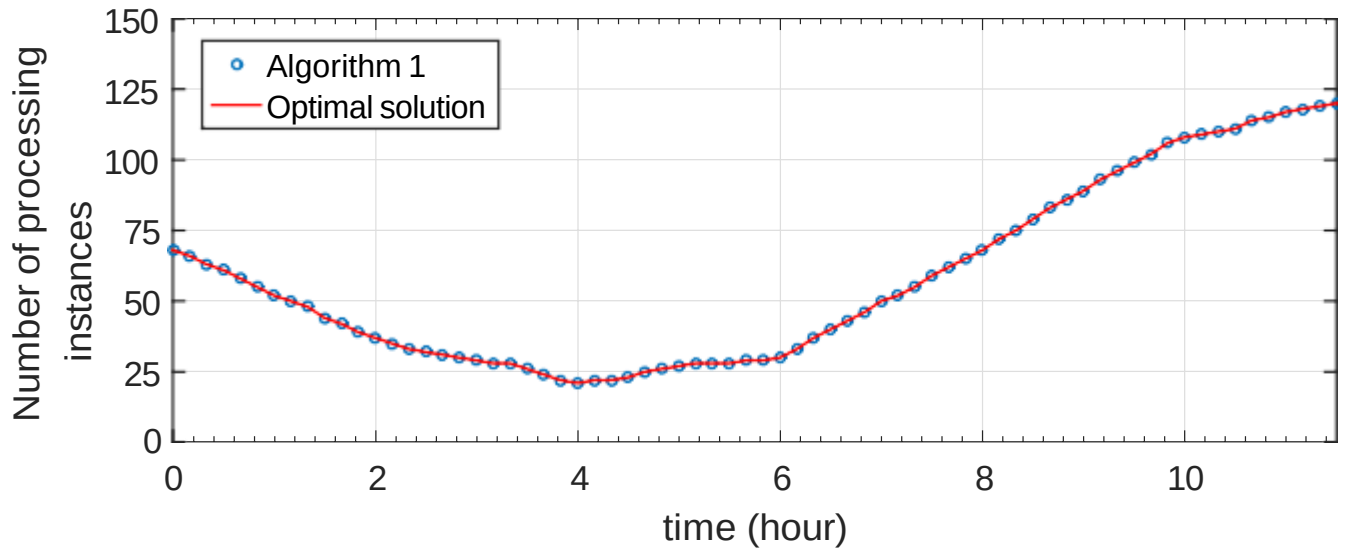

Fig. 6: Comparison of the required computational resources estimated by our proposed dimensioning algorithm and the optimal solution.

TABLE II: Parameters Configuration

| | |
|---|---|
| External arrival process at the GO & Workload Predictor | |
| External arrival process at the GO | Modulated Poisson process whose arrival rate versus time is given in Fig. 5 |
| Workload predictor | Focused time delay neural network with 10 neurons |
| Root mean squared error (RMSE) of the workload predictor | 361.5 SORs per second |
| Service processes | |
| Service rate of each processing instance | 10000 packets per second |
| SCV of the processing instance service time | 0.65 |
| QoS requirements | |
| $T_{max}$ | $2\ ms$ |

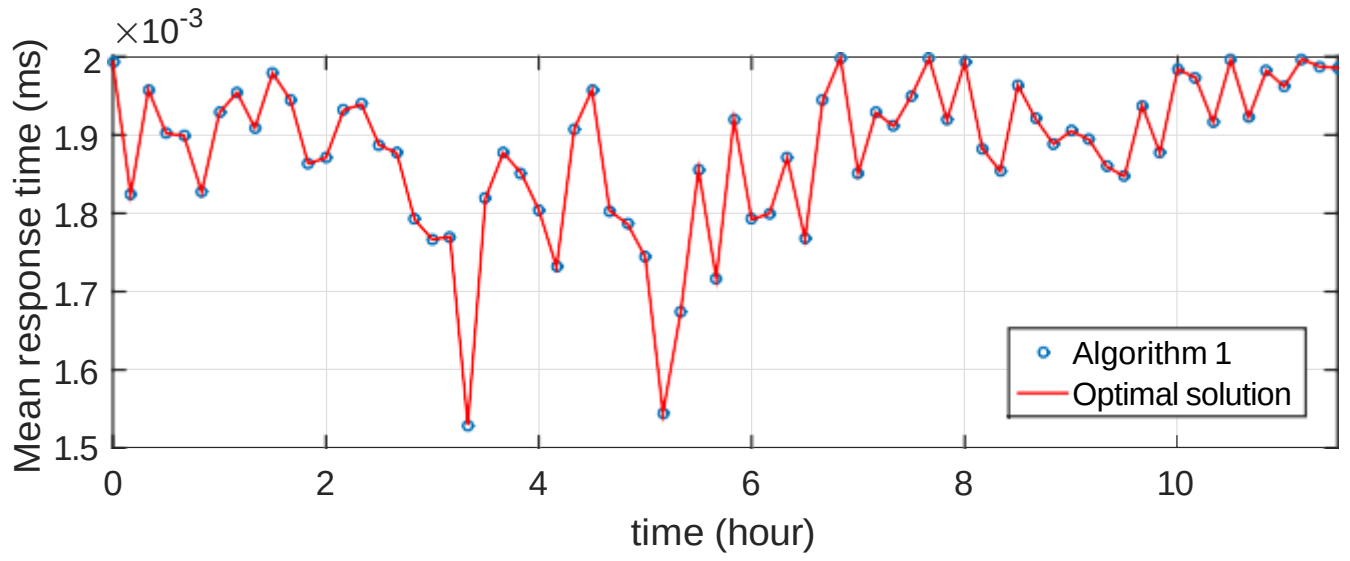

Fig. 7: Comparison of the overall mean response time of the system achieved by our algorithm and the optimal solution.

The simulator also includes an implementation of our DRP algorithm. As workload predictor, we used the focused time delay neural network model with a tapped delay line with a maximum delay of 2 and ten neurons in the hidden layer. To train the neural network, we used the temporal distribution of the workload for a four weeks period included in [24] (see [24, Fig.1]). We chose Levenberg-Marquardt as training algorithm. Figure 5 shows the relative error of the workload predictor. The decisions to provision are taken periodically every ten minutes. A single token bucket was used as request policer at the input of the GO. The tokens generation rate depends on the resources allocated to the NSOS. More precisely, it is computed as the maximum mean external arrival rate $\lambda_{0GO}$ given the current capacity of the NSOS using the model presented in Section V. The reactive provisioning mechanism is triggered either when $(\lambda_{0GO} - \lambda_{0GO,pred})/\lambda_{0GO} \geq 0.05$ or $(\lambda_{0GO} - \lambda_{0GO,pred})/\lambda_{0GO} \leq 0.5$. The VM instantiation time was set to 82 seconds [47]. The target mean response time of the system to serve a slice orchestration request was set to 2 ms, i.e., $T_{max} = 2\,ms$.

## *B. Dimensioning Algorithm Time Complexity and Optimality*

In order to gauge the performance of the proposed resources dimensioning algorithm for the NSOS (i.e., the Algorithm 1), we considered three metrics: i) the difference between the resources estimated by the algorithm and the optimal solution, ii) the difference between the mean response time achieved by the algorithm and the optimal solution, and iii) the time complexity of the algorithm. We computed the optimal solution of the problem using the exhaustive search method described in Section IV.

Figures 6 and 7 depict respectively the comparison between our algorithm and the optimal solution to estimate the amount of required computational resources and the mean response time achieved with each estimation. Since finding the optimal solution is very computationally intensive, the scenario considered in this evaluation had only one DSO entity and the evaluation was carried out only for the first twelve hours of the day. As it is observed, our solution achieves the optimality goals for the scenario considered. Moreover, the same test was conducted trying different combinations for the setup of the input parameters. Specifically the following ranges of values were sampled for the input parameters: $0 \leq c_{a0}^2 \leq 10$, $0 \leq c_{se}^2 \leq 10$, $\sum_{e \in E} V_e \cdot 1/\mu_e < T_{max} \leq 10ms$, and $0 \leq c_{se}^2 \leq 10$, and $1 \leq m_{max} \leq 15$. The algorithm found the optimal solution in all cases tested.

This result can be explained if the performance model described in Section V is a convex and non-increasing sequence with the number of CPU cores allocated to each entity. In such case, the strategy followed by Algorithm 1 will find a solution where $T \leq T_{max}$ with the least number of iterations possible or, equivalently, minimizing the total number of allocated CPU cores (which is the objective). For non-convex performance models, it is not assured that Algorithm 1 will find the optimal solution.

Finally, we studied the dependence of the execution time of our algorithm on its input parameters (see Figs. 8a and 8b). Each point in Figs. 8a and 8b represents the average of the measurements obtained for ten independent runs.

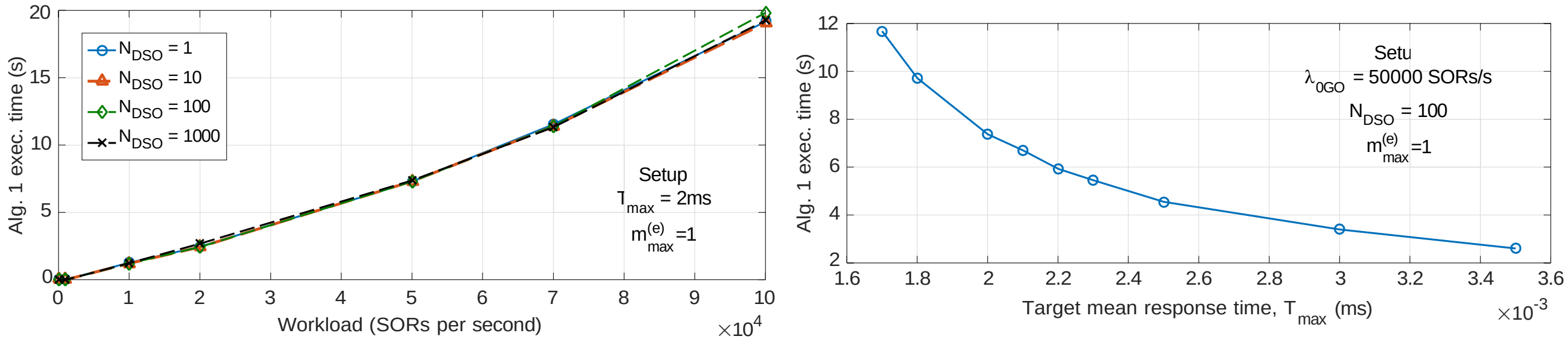

(a) (b)

Fig. 8: Time complexity study for the resources dimensioning algorithm (Algorithm 1): a) Execution time versus the workload and number of NSOS components. b) Execution time versus the overall target mean response time.

Figure 8a shows the algorithm execution time versus the workload for different number of DSOs $N_{DSO}$. Observe that each additional DSO included in the scenario accounts for six additional network entities in the system (i.e., DSO, DSSDNC, DSNFVO, DSVIM, DSRRO, DSeNBs). Then, for $N_{DSO} = 2$ the slices orchestration system is composed of fifteen different entities ($N_e = 15$). As shown in Fig. 8a the algorithm has quadratic time complexity with the workload $\lambda_{0GO}$. Specifically, the shape of the function

$$f(\lambda_{GO_0}) = 8.9 \cdot 10^{-10} \cdot \lambda_{0GO}^{2} + 10^{-4} \cdot \lambda_{0GO} + 6.3 \cdot 10^{-3}$$

fits the experimental curve for $N_{DSO} = 1$ with an R-squared of 97.5%. The rationale behind this result is that our algorithm requires $M^{*} - M_0$ evaluations of the performance model which, in turn, has a time complexity $O(M)$. The result is also obtained analytically in Appendix **??** for the case $N_e = 1$.

It is also shown in Fig. 8a that the execution time of the Algorithm 1 does not depend on the number of entities $N_e$. As pointed out, the NSOS performance model detailed in Section V has a linear time complexity with the number of queuing nodes $K$ each of which represents an instance of a given NSOS entity. Although for low workloads the total number of instances is given by the number of entities (at least we will have one instance per entity), for high workloads the number of queues are dominated by the workload. Figure 8b depicts impact of the target mean response time $T_{max}$ on the algorithm execution time. In contrast to the analysis provided in Appendix **??**, the experimental results show that the algorithm has a logarithmic time complexity with the inverse of $T_{max}$, i.e., $O(\log T_{amx}^{-1})$. The model

$$f(T_{max}) = a \cdot \log\left(T_{max} - \sum_{e\in E} V_e \cdot \frac{1}{\mu_e}\right)^{-1} + b \qquad (26)$$

fits the experimental curve with an R-squared of 95.5% when $a = 3.2$ and $b = -17$.

In summary, the experimental results suggest that Algorithm 1 has time complexity $O\left(\lambda_{0GO}^{2} \cdot \log T_{max}^{-1}\right)$ when $M^{*} < M_{max}$.

### C. *Dynamic Provisioning*

Finally, we checked that our solution works properly by means of simulations. In this assessment, we considered $N_{DSO} = 3$. The simulations were repeated ten times.

Fig. 9 depicts the total required number of processing instances over time predicted by our DRP algorithm. In the same way, Fig. 10 shows the number of processing instances allocated to each entity over time according to our DRP solution. As the GO has to process the highest number of messages per control procedure, it presents the greatest demand of resources. As shown in Fig. 11, for the resources allocation performed by our DRP solution the maximum delay threshold is always met, thus validating the proper operation of it. Please note that Fig. 11 includes the average (solid line in blue) and the 95% confidence interval (shaded area in red) of the simulation results.

Figure 12 shows the rejection rate of the NSOS request policing mechanism, i.e., the percentage of the slice orchestration requests discarded per unit time. The maximum 95% confidence interval obtained for this curve was 0.21%. Interestingly, the rejection rate is greater than 5% at some points, even though we used $(\lambda_{0GO} - \lambda_{0GO,pred})/\lambda_{0GO} \geq 0.05$ as a condition to trigger the reactive provisioning mechanism. This can be explained by the fact that we considered a high VM instantiation time (82 s). We observed that a maximum rejection rate of 5% is only attained considering no VM instantiation time. Please note that the highest values of the rejection rate are between 6 and 10 hours. During this period, the workload predictor exhibits the highest prediction error underestimating the foreseen workload.

In summary, the above results suggest that it is not feasible to guarantee an instantaneous maximum rejection rate below a given threshold. Instead, we can only guarantee a maximum rejection time during a time interval of a given duration. This performance metric will be determined by the reaction time of the reactive provisioning mechanism which, in turn, strongly depends on the VM instantiation time.

## VIII. Conclusion

In this work, we introduce a hierarchical architecture that could be used to practically design and deploy a scalable E2E multi-domain mobile network slicing orchestration system in an efficient manner. We model the interactions between the components that make up the NSOS using queuing theory and validated our model with a viable system simulation. In addition, we showcase a dynamic auto-scaling algorithm that could be used to enable the autonomous operation of the system. Our auto-scaling algorithm efficiently scales the resources of the multi-domain orchestration system while maintaining a system-wide stability and consistently achieving the E2E multi-domain network slice creation requests based on defined requirements. Ultimately, the system maximizes the orchestration of E2E global network slices based on the available system resources and set orchestration time policy. In subsequent works, we plan to include further detailed conditions that could influence the functionality and performance of the orchestration system.

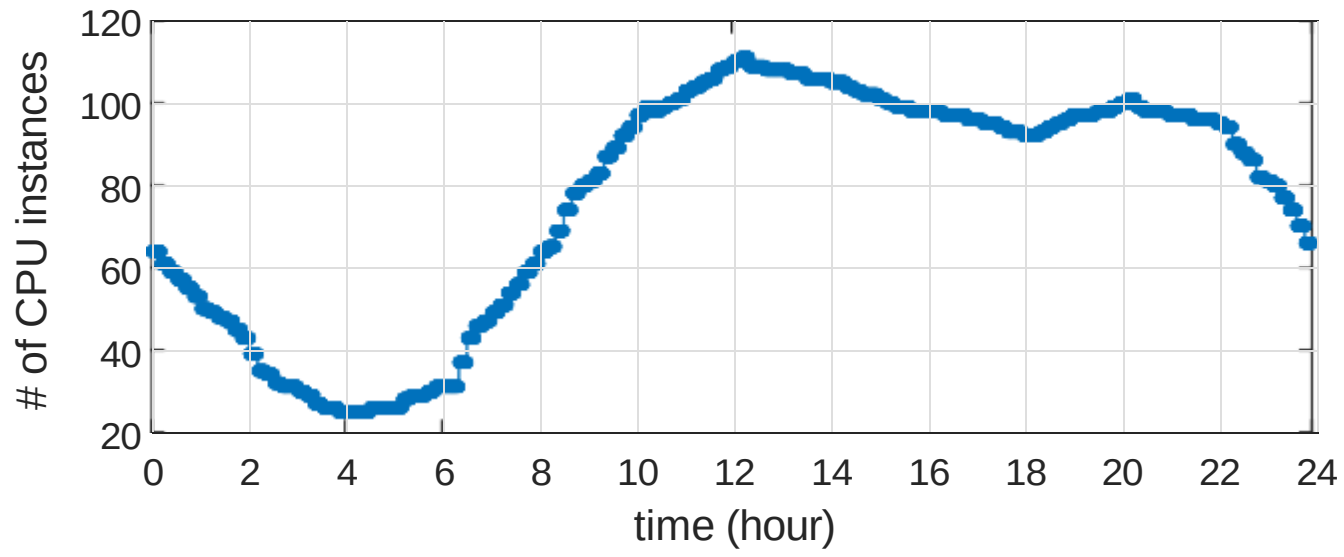

Fig. 9: Total number of processing instances required by the Slice Orchestrator System to met the delay budget (2 ms).

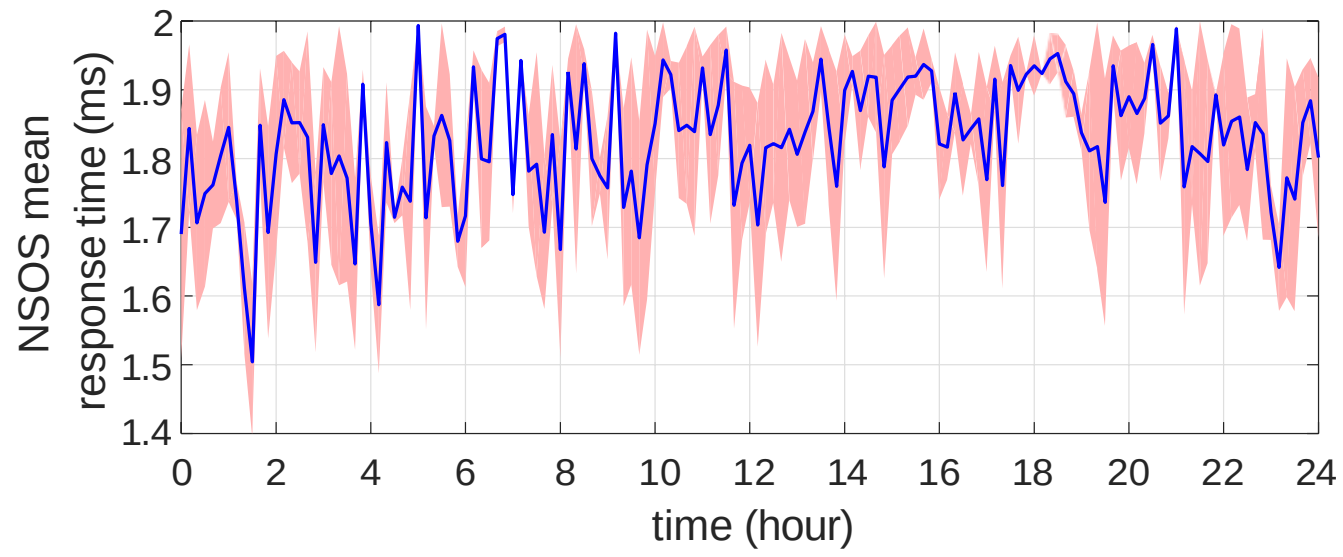

Fig. 11: Slices Orchestration System response time.

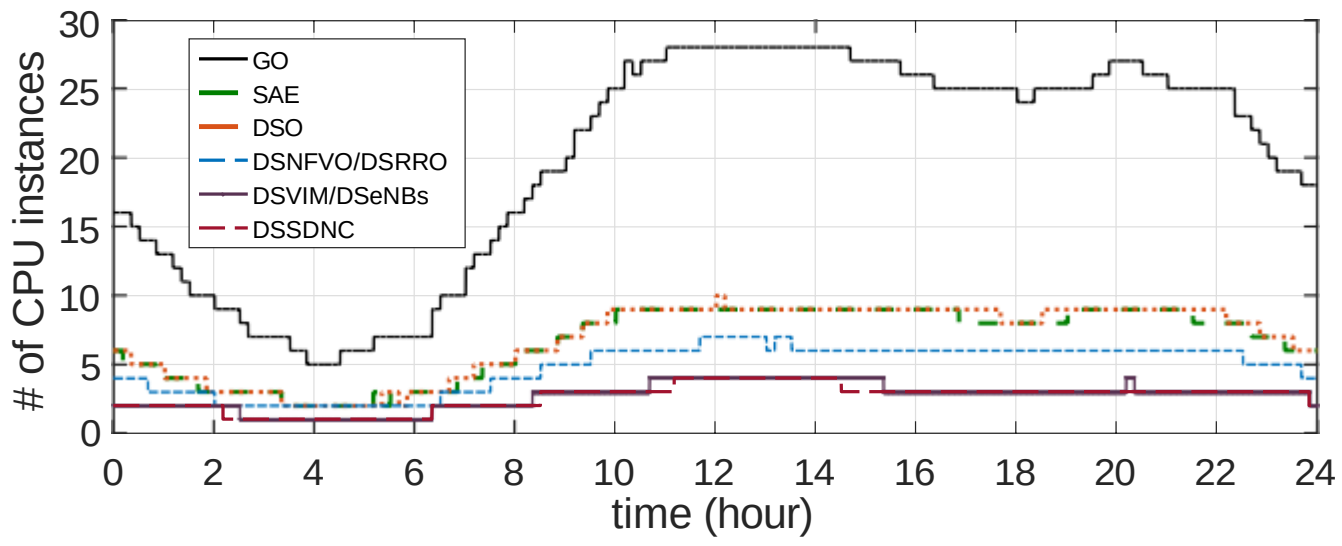

Fig. 10: Number of processing instances allocated to each entity to met the delay budget (2 ms).

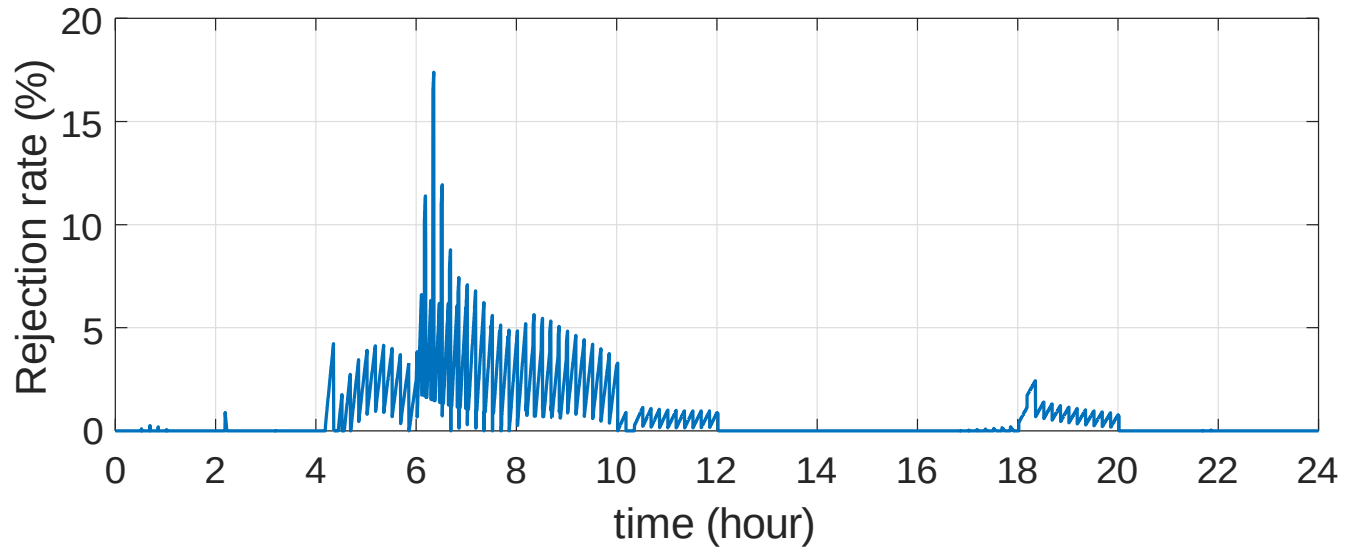

Fig. 12: Rejection rate of the network slice orchestration requests.

## Acknowledgment

The project is partially supported by the European Union's Horizon 2020 research and innovation program under the 5G!Pagoda project, the MATILDA project and the 6Genesis project with grant agreement No. 723172, No. 761898 and No. 318927 respectively and also was partially funded by the Academy of Finland Project CSN - under Grant Agreement 311654 and the Spanish Ministry of Education, Culture and Sport (FPU Grant 13/04833), and the Spanish Ministry of Economy and Competitiveness and the European Regional Development Fund (TEC2016-76795-C6-4-R).

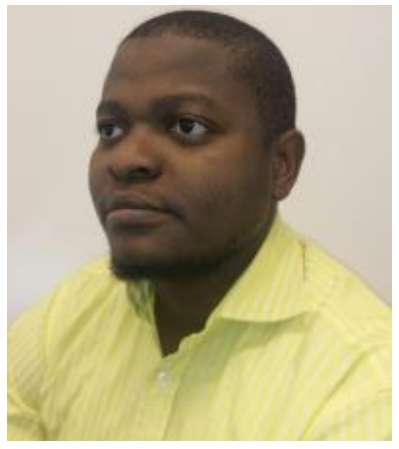

**Ibrahim Afolabi** obtained his Bachelors degree from VAMK University of Applied Sciences, Vaasa, Finland, in 2013 and his Masters degree from the School of Electrical Engineering, Aalto University, Finland in 2017. He is presently pursuing his doctoral degree at the same university where he obtained his Masters degree from and his research areas include Network Slicing, MEC, network softwerization, NFV, SDN, and dynamic network resource allocation.

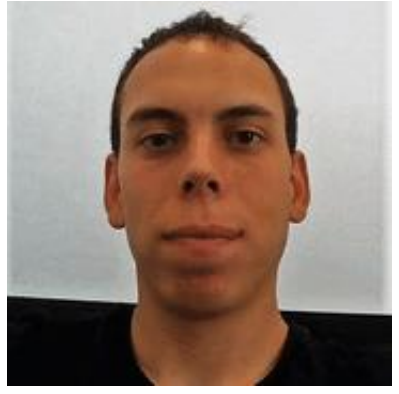

**Jonathan Prados-Garzon** received his B.Sc., M.Sc., and Ph.D. degrees from the University of Granada (UGR), Granada, Spain, in 2011, 2012, and 2018, respectively. Currently, he is a post-doc researcher at MOSA!C Lab, headed by Prof. Tarik Taleb, and the Department of Communications and Networking of Aalto University, Finland. His research interests include Mobile Broadband Networks, Network Softwarization, and Network Performance Modeling.

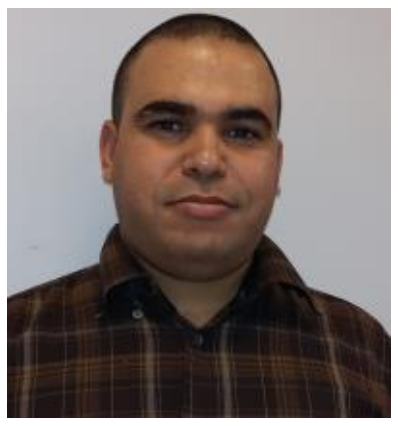

**Miloud Bagaa** received the bachelors, masters, and Ph.D. degrees from the University of Science and Technology Houari Boumediene Algiers, Algeria, in 2005, 2008, and 2014, respectively. He is currently a Senior Researcher with the Communications and Networking Department, Aalto University. His research interests include wireless sensor networks, the Internet of Things, 5G wireless communication, security, and networking modeling.

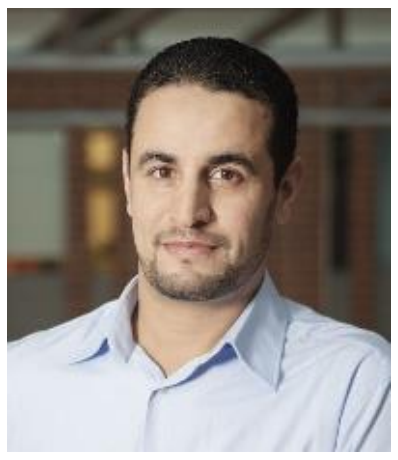

**Tarik Taleb** received the B.E. degree (with distinction) in information engineering in 2001, and the M.Sc. and Ph.D. degrees in information sciences from Tohoku University, Sendai, Japan, in 2003, and 2005, respectively. He is currently a Professor with the School of Electrical Engineering, Aalto University, Espoo, Finland. He is the founder and the Director of the MOSA!C Lab. He is the Guest Editor-in-Chief for the IEEE JSAC series on network Softwarization and enablers.

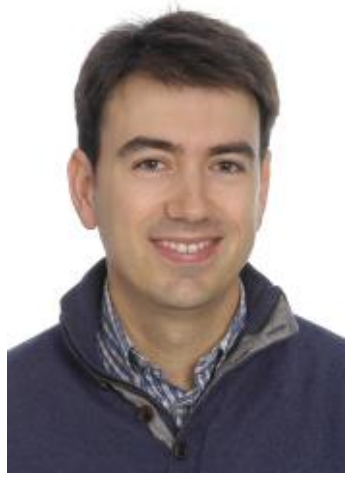

**Pablo Ameigeiras** received his M.Sc.E.E. degree in 1999 from the University of Malaga, Spain. He carried out his Masters thesis at the Chair of Communication Networks, Aachen University, Germany. In 2000 he joined Aalborg University, Denmark, where he carried out his Ph.D. thesis. In 2006 he joined the University of Granada, where he has led several projects in the field of LTE and LTE Advanced systems. Currently his research interests include the application of the SDN/NFV paradigms for 5G systems.